\documentclass[11pt,twoside]{article}
\usepackage{asp2006}
\usepackage{txfonts}
\usepackage{graphicx}
\begin{document}
\title{Deep wide-field GMRT surveys at 610~MHz}
\author{D.~A.~Green$^1$ \& T.~S.~Garn$^{1,2}$}
\affil{$^1$Cavendish Laboratory,
      19 J.~J.~Thompson Avenue,
      Cambridge,
      CB3~0HE, United Kingdom}
\affil{$^2$SUPA, Institute for Astronomy,
       Royal Observatory Edinburgh,
       Blackford Hill,
       Edinburgh,
       EH9~3HJ, United Kingdom}

\begin{abstract}
The GMRT has been used to make deep, wide-field surveys of several fields at
610~MHz, with a resolution of about 5~arcsec. These include the Spitzer
Extragalactic First Look Survey field, where 4 square degrees were observed
with a r.m.s.\ sensitivity of about 30~$\muup$Jy~beam$^{-1}$, and several SWIRE
fields (namely the Lockman Hole, ELAIS-N1 and N2 fields) covering more than 20
square degrees with a sensitivity of about 80~$\muup$Jy~beam$^{-1}$ or better.
The analysis of these observations, and some of the science results are
described.
\end{abstract}

%------------------------------------------------------------------------------%
\section{Introduction}

In recent years we have used the GMRT (e.g.\ \citealt{2002IAUS..199..439R}) to
make relatively deep, wide area observations of several fields at 610~MHz, with
a resolution of about 5~arcsec. The deepest of these fields is the Spitzer
`Extragalactic First Look Survey' (xFLS) region\footnote{See: {\tt
http://ssc.spitzer.caltech.edu/fls/}}, which is about 4 square degrees in
extent, and is in a direction with very little Galactic infra-red foreground
emission. This field already has deep VLA observations at 1.4~GHz available
with a resolution of about 5~arcsec (see \citealt{2003AJ....125.2411C}), for
which the GMRT provides complementary 610-MHz observations at a similar
resolution. The other fields are selected areas from the `SWIRE' ({\bf S}pitzer
{\bf W}ide-area {\bf I}nfra{\bf R}ed {\bf E}xtragalactic) survey
(\citealt{2003PASP..115..897L}). These areas have deep infra-red observations
from the Spitzer Space Telescope, and often other deep complementary optical
observations, but do not have deep, wide-field radio observations available.

Table~\ref{tab:surveys} gives a summary of the xFLS and SWIRE fields which were
observed.  Given the primary beam of the GMRT at 610~MHz is $\approx
43$~arcmin, each pointing covers $\approx 0.4$~degree$^2$. For the ELAIS-N1
field, four of the nineteen pointings (which overlap regions where deeper
optical observations are available; \citealt{astro-ph/0703037}) were observed
more deeply, to provide a noise value of $\approx 40$~$\muup$Jy~beam$^{-1}$,
rather than the $\approx 70$~$\muup$Jy~beam$^{-1}$ of the other pointings in
this field. Also, observations of an additional 26 outer pointings in the
Lockman Hole field have been obtained -- in order to cover the whole $\approx
11$~degree$^2$ observed with Spitzer -- and analysis of these is ongoing. All
of the observations were made with two 16-MHz sidebands, in both right and left
circular polarised emission. Each sideband was divided into 128 spectral
channels, so that any narrow band interference could be identified and removed.

\begin{table}
\caption{Fields observed with the GMRT at 610~MHz.\label{tab:surveys}}
\smallskip
\centering
\small
\begin{tabular}{lcccl}
\tableline
\noalign{\smallskip}
field        & pointings &  noise  & sources    & reference                  \\
             &           & ($\muup$Jy~beam$^{-1}$) & &                       \\
\noalign{\smallskip}
\tableline
\noalign{\smallskip}
xFLS         &     7     &    27   &   3944     & \cite{2007MNRAS.376.1251G} \\
ELAIS-N1     &    19     &  40/70  &   2500     & \cite{2008MNRAS.383...75G} \\
Lockman Hole &    12     &    60   &   2845     & \cite{2008MNRAS.387.1037G} \\
ELAIS-N2     &    13     &    80   & $\sim$1500 & in preparation             \\
\noalign{\smallskip}
\tableline
\end{tabular}
\end{table}

%------------------------------------------------------------------------------%
\section{Data Analysis}

The analysis of these observations was performed using classic {\sc AIPS}.
Initial careful editing of the data was necessary in order to remove
interference, and other bad data (e.g.\ a few systematically poor baselines,
due to correlator hardware problems). Bright `primary' calibrators (e.g.\ 3C286
and/or 3C48) were observed at the start/end of each observation run, in order
to define the flux density scale, and also derive antenna-based bandpass
corrections. The bandpass corrections were applied, and several central
channels in each sideband were averaged together, in order to make a
`pseudo-continuum' channel (`channel 0' in VLA notation). The antenna-based
amplitude and phase calibrations of the telescope were derived using the
pseudo-continuum channel, from the observations of nearby, compact `secondary'
calibrators, which were observed every 30 min or so. The antenna-based
amplitude/phase and bandpass corrections were then applied to the observed
visibilities, and 10 channels were averaged together, in order to reduce the
size of the $uv$ datasets before imaging.

The wide field-of-view of the GMRT means that multiple small `facets' need to
imaged, and then combined together. Generally 19 facets were used (i.e.\ a
central facet, 6 in a surrounding ring, and 12 in a larger ring on a hexagonal
grid), to cover the full observed field. The quality of the synthesised images
were improved by several cycles of self-calibration, at 10, 3 and 1 min for
phase only, and finally 10 min for amplitude and phase. This lead to images
with dynamic ranges -- peak to r.m.s.\ away from bright sources -- of several
thousand to one, with r.m.s.\ noise values away from bright sources that are
close to the expected thermal noise. Near bright sources, however, the quality
of the images is limited, which is believed to be due to variations in the
pointing of the telescope antennas (see below).

The early analyses of the overlapping pointings from the xFLS field revealed
two problems: (1) the time stamps of $uv$-data were slightly wrong (by about
7~s), meaning that the $uv$ coordinates were also slightly wrong, leading to a
distortion (nearly a rotation) in the synthesised images; (2) comparison of the
flux densities of compact sources in overlap regions between adjacent pointings
showed systematic differences, which are thought to be due to antenna pointing
offsets during the observations, particularly at low elevations. For the former
problem, an {\sc AIPS} task was written\footnote{See: {\tt
http://www.mrao.cam.ac.uk/{\char'176}dag/UVFXT/}} to correct affected data, by
adjusting the time stamps and recalculating the $uv$ coordinates (and the
real-time GMRT system was corrected in the summer of 2006). The latter problem
was correctly empirically, using a slightly shifted effective primary beam
position.  An improved pointing model is currently being implemented at the
GMRT to address this problem.

Further details of the analysis procedures used are given in
\cite{2007MNRAS.376.1251G} and \cite{GarnPhD}. Users of classic {\sc AIPS}
should also note that there were problems with the tasks {\sc SPLIT} and {\sc
SPLAT} which might result in incorrect frequency headers when averaging
spectral channels together. These tasks were patched late in 2007\footnote{See:
patches 17, 18 and 19 at {\tt http://www.aips.nrao.edu/31DEC06/patches.html}}.

%------------------------------------------------------------------------------%
\section{Results and Discussion}

\begin{figure}
\centerline{\includegraphics[width=13.5cm]{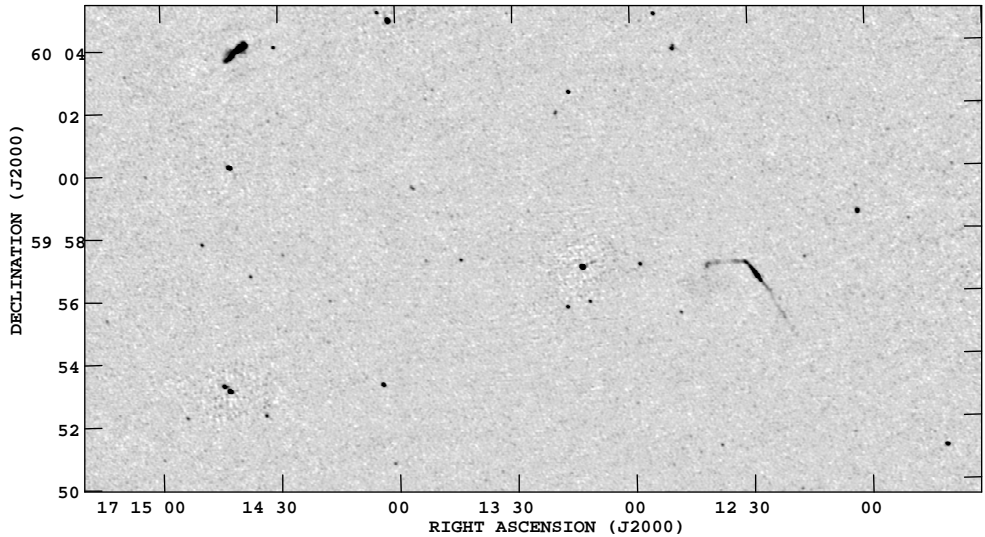}}
\caption{A portion of the xFLS region observed at 610~MHz with the GMRT.  The
grey-scale ranges between $-0.2$ and 1~mJy~beam$^{-1}$, and the resolution is
$5.8\times4.7$~arcsec$^{2}$, at a position angle
$+60^{\circ}$.\label{fig:xFLS}}
\end{figure}

Given the wide field of the 610-MHz images produced by these observations, it
is not possible to show these in full (e.g.\ the xFLS field is $\approx
2$~degree across, so with 1.5~arcsec pixels to well sample the $\approx
5$~arcsec beam, the image is about 5000 pixels in each dimension).
Figure~\ref{fig:xFLS} shows a small portion -- only about 3~per~cent -- of the
deep GMRT image of the xFLS field.

The GMRT results for the xFLS field are comparable in resolution, and in depth
for synchrotron emission, to the available deep VLA 1.4-GHz survey of the same
region: the GMRT observations, which took about 40 hours of observations, have
an r.m.s.\ noise (but only away from the bright sources) of $\approx
30$~$\muup$Jy~beam$^{-1}$; the VLA observations, which took about 200 hours of
observations, have an r.m.s.\ noise of $\approx 23$~$\muup$Jy~beam$^{-1}$.

These deep, wide-field surveys provide catalogues of radio sources, which then
allow a variety of astrophysical studies to be made, including comparisons with
the deep infra-red and other wavelength observations available for these
fields. In particular, for the xFLS region, the existence of both the deep VLA
1.4-GHz and GMRT 610-MHz observations allow the properties of many sub-mJy
radio sources -- which are dominated by star forming galaxies, rather than AGN
-- to be studied, including their radio spectral indices. Some of these studies
are described briefly below.

In the xFLS survey region it is possible to define a sample of 235 sources
which are detected at both 610 and 1410~MHz in the radio and at 24 and 70
$\muup$m in the infra-red, for which spectroscopic redshifts (up to $z \approx
1$) are available. From these data it is possible to study the infra-red/radio
correlation, applying `$k$-corrections' to provide rest-frame luminosities at
the same frequencies (using the observed spectral indices in the radio, and
model SEDs in the infra-red), see \cite{2007mru..confE..73G}. Any cosmic
evolution of the magnetic fields in galaxies would be expected to be reflected
in changes in the comparative radio brightness of star-forming galaxies. The
available data shows no evidence for such an effect, suggesting that there has
been little evolution in the magnetic fields of galaxies since $z \sim 1$.

\begin{figure}
\centerline{\includegraphics[angle=270,width=5.5cm]{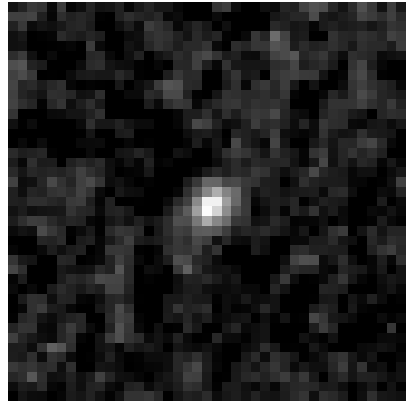}}
\caption{A median 610-MHz stacked image, made from radio images that were
centred on the locations of 591 sources with 24-$\muup$m flux density between
150 and 210~$\muup$Jy -- the grey-scale range is $-2$ to $+20$~$\muup$Jy
beam$^{-1}$.\label{fig:stack}}
\end{figure}

Although the GMRT observations provide deep images, it is possible to probe,
statistically, the properties of sources that are too faint to be detected
individually, using a median `stacking' technique (see
\citealt{2008arXiv0812.0281G}). For example, stacking 591 sources from the xFLS
field with Spitzer 24-$\muup$m flux densities between 150 -- 201~$\muup$Jy
leads to a median detected radio flux of $21 \pm 2$~$\muup$Jy at 610~MHz, see
Figure~\ref{fig:stack}. Using this technique, the radio properties of sources
with 24-$\muup$m flux densities between 150 $\muup$Jy and 2~mJy show no
evidence for a change in the infra-red/radio correlation.

\begin{figure}
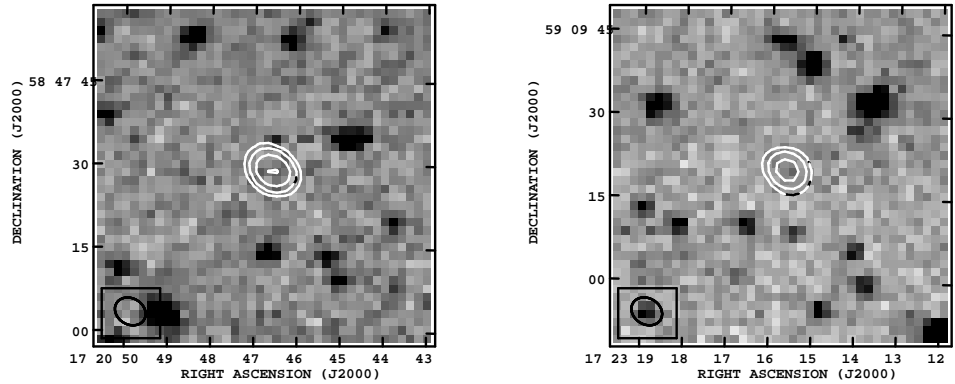

\centerline{\includegraphics[width=6.0cm]{green3a} \qquad
            \includegraphics[width=6.0cm]{green3b}}
\caption{Examples of two radio sources (610-MHz contours of 0.5, 1, 2 and
4~mJy) which are not detected at 3.6~$\muup$m (grey-scale at $\pm
1\sigma$).\label{fig:irfs}}
\end{figure}

Almost all of the radio sources detected in these GMRT surveys are also
detected in the infra-red due to the sensitivity of the Spitzer surveys, as is
expected for local AGN or star-forming galaxies. However, there are a small
number of relatively bright radio sources that are not detected in the
infra-red \citep{2008MNRAS.391.1000G}. Two examples of such sources are shown
in Figure~\ref{fig:irfs}. There are 14 of these infrared-faint radio sources in
the xFLS field, 8 of which are detected near to the limiting magnitude (24.5)
of a deep R-band image of the region. These sources can modelled as compact
($<20$~kpc) Fanaroff--Riley type II radio galaxies located at high redshift ($z
\sim 4$).

These surveys have also been used to construct differential source counts at
610 MHz between 270~$\muup$Jy and 200~mJy (\citealt{2008MNRAS.387.1037G}). A
flattening in the differential source counts is clearly visible below 2~mJy,
and a three-component population containing steep and flat spectrum AGN, and
starburst galaxies which undergo pure luminosity evolution is sufficient to
model simultaneously both the 610-MHz source counts, and 1.4-GHz source counts
from the literature.

%------------------------------------------------------------------------------%
\acknowledgements
We thank the staff of the GMRT that made these observations possible. GMRT is
run by the National Centre for Radio Astrophysics of the Tata Institute of
Fundamental Research.

%------------------------------------------------------------------------------%

%------------------------------------------------------------------------------%

\begin{thebibliography}{}

\bibitem[Condon et al.(2003)]{2003AJ....125.2411C}
  Condon, J.~J., Cotton,  W.~D., Yin, Q.~F., Shupe, D.~L.,
  Storrie-Lombardi, L.~J., Helou, G., Soifer, B.~T., \& Werner, M.~W.\
    2003, AJ, 125, 2411

\bibitem[Garn(2009)]{GarnPhD}
  Garn, T.~S.
    2009, PhD thesis (University of Cambridge, UK)

\bibitem[Garn \& Alexander(2008)]{2008MNRAS.391.1000G}
  Garn, T., \& Alexander, P.\
    2008, MNRAS, 391, 1000

\bibitem[Garn \& Alexander(2009)]{2008arXiv0812.0281G}
  Garn, T., \& Alexander, P.\
    2009, MNRAS, in press (arXiv:astro-ph/0812.0281)

\bibitem[Garn et al.(2007a)]{2007mru..confE..73G}
  Garn, T., Ford, D.,  Alexander, P., Green, D.~A., \& Riley, J.~M.\
    2007a, in The Modern Radio Universe: From Planets to Dark Energy,
    Proceedings of Science,
    ({\tt http://pos.sissa.it/contribution?id=PoS(MRU)073})

\bibitem[Garn et al.(2007b)]{2007MNRAS.376.1251G}
  Garn, T., Green, D.~A., Hales, S.~E.~G., Riley, J.~M., \& Alexander, P.\
    2007b, MNRAS, 376, 1251

\bibitem[Garn et al.(2008a)]{2008MNRAS.383...75G}
  Garn, T., Green, D.~A., Riley, J.~M., \& Alexander, P.\
    2008a, MNRAS, 383, 75

\bibitem[Garn et al.(2008b)]{2008MNRAS.387.1037G}
  Garn, T., Green, D.~A., Riley, J.~M., \& Alexander, P.\
    2008b, MNRAS, 387, 1037

\bibitem[Lonsdale et al.(2003)]{2003PASP..115..897L}
  Lonsdale, C.~J., et al.\
    2003, PASP, 115, 897

\bibitem[Pramesh Rao(2002)]{2002IAUS..199..439R}
  Pramesh Rao, A.\
    2002, in IAU Symposium 199, The Universe at Low Radio Frequencies,
     ed. A. Pramesh Rao, G. Swarup \& Gopal Krishna (San Fransico: ASP),
     439

\bibitem[Warren et al.(2007)]{astro-ph/0703037}
  Warren, S.~P., et al.\
    2007, arXiv:astro-ph/0703037

\end{thebibliography}
\end{document}